\documentstyle[12pt,epsf]{article}
\begin{document}

\parskip=0pt
\parindent=125mm
 hep-ph/9602310\par
TECHNION-PH-96-5\par
 February, 1996\par

\bigskip
\bigskip
\begin{center}
\parskip=10pt

{\huge Proton lifetime, Yukawa couplings and dynamical SUSY breaking 
in SU(5) GUT}
\bigskip
\bigskip

{\large  Boris Blok, Cai-Dian L\"u and Da-Xin Zhang }

 Physics Department, Technion- Israel Institute of Technology,

 Haifa 32000, Israel

\end{center}

\parindent=23pt
\bigskip
\begin{abstract}
We study the influence of messenger Yukawa couplings and 
top, bottom and $\tau$ Yukawa couplings on the proton lifetime 
in SU(5) Supersymmetric GUT
  with dynamical supersymmetry breaking mechanism due to Dine and Nelson.
\end{abstract}

\newpage
\parskip=5pt

 \section{Introduction.}

\par Recently, there was a great increase of interest in the study of the 
physics beyond the standard model, in particularly in the
study of the supersymmetric grand 
unification theories (SUSY GUTs).
This interest was in particular stimulated by the new  data 
due to LEP1 and LEP1.5  experiments. If one  believes literally in all 
this data, especially in the $R_b$ measurements, one comes to the conclusion
that neither  the standard model (SM) nor  
the minimal supersymmetric standard model
(MSSM) are  sufficient to explain the experimental results \cite{1,2,3}. 
 One naturally has to turn to the
 other models. One of the most popular directions in the search 
 for the new physics is the study of SUSY GUTs.

Unfortunately, one immediately encounters the problem of how to break 
supersymmetry (SUSY). The most popular way is to use a hidden sector through 
which SUSY breaking is transferred to 
the visible world by means of supergravity
 ( See e.g. ref.\cite{4} for a review). 
Recently a new approach to  break SUSY was developed\cite{5,6,7}. In this 
approach 
 SUSY  is broken dynamically and  the effects of SUSY
breaking are transferred from hidden sector  to the visible world
by means of gauge interactions. 
The natural question is whether one can 
use this mechanism to construct realistic SUSY GUT theories \cite{8}. 

There are several tools one can use at present to decide whether the GUT
theory is realistic or not. One of the most powerful among them
 is the experimental limit
 on the proton lifetime.
It was shown in ref.\cite{8} that simple extension of mimimal
SU(5) SUSY GUT with the mechanism of dynamical SUSY breaking
due to refs. \cite{5,6,7} 
leads to the too  small lifetime of  proton. 
Consequently, this model is ruled out.

The  purpose of this note is to
return once again to the analysis of the proton lifetime in the minimal 
SU(5) SUSY GUT with dynamical supersymmetry breaking due to Dine and Nelson
, and to
 include the Yukawa couplings in the 
analysis of ref. \cite{8}.
  We study the influence  of Yukawa couplings,
both of the 3rd generation and of the hiden sector, on the proton lifetime
and masses of the color triplet Higgs boson $M_{H_c}$. 
We shall see that the  
inclusion of Yukawa couplings increases
the proton lifetime but not significantly. 
However, the model cannot be ruled out at $2\sigma$ level. The model is 
not ruled out
in a small window  at small $\tan \beta$
if one takes into account the uncertainties in the LEP data for
 $\alpha_1(m_Z)$ and $\alpha_2(m_Z)$ $-$ QED and weak coupling constants at 
$m_Z$ scale.

 The note is organised in the following way:
 First we use renormalization group (RG) equations for gauge couplings
 taking  Yukawa couplings into account. 
 We then find the masses of color triplet
 bosons $M_{H_c}$, depending on $\alpha_3(m_Z)$ (QCD coupling) 
 for different values of $\alpha_1$, $\alpha_2$. 
 We then use the
 formula for proton lifetime from ref.\cite{13} to find the proton lifetime
 for different allowed by LEP data
 values of $\alpha_1(m_Z)$ and $\alpha_2(m_Z)$,
 $\alpha_3(m_Z)$. 
 Our final results are depicted in figs.1-6, thus giving the bounds 
 on the possibility of using mechanism of refs.\cite{5,6,7} to break 
 SUSY in SU(5) GUT model. 

\section{Calculation.}

Let us briefly describe the model we are going to study (we refer 
to the reader to refs.\cite{5,6,7,8} for details).
 
 The model is the mimimal SU(5) SUSY  GUT  plus the messenger superpotential
\begin{equation}
\Delta W_m = \lambda_L S \bar L L + \lambda_D S \bar D D.
\end{equation}
Here L and D, that carry   quantum numbers of SM gauge groups
 and form together into $5 +\bar 5$ of SU(5), 
are the  messenger lepton and quark chiral superfields added to the model.
S is a  scalar chiral superfield singlet under the SM gauge groups.
S transfers
the supersymmetry breaking from hidden sector to the fields of the standard
model. 
At the messenger scale $\Lambda_m \sim 100$ TeV 
the singlet S and its F-component
get vacuum 
expectation values and SUSY breaks down. 
The  heavy superfields L and  D transfer the information of 
SUSY breaking to the fields 
of SM through their couplings to the SM gauge groups.
The
  analysis of proton decays
 in this note  is similiar to the work by Carone and 
Murayama\cite{8}. 
However,  these authors did not take into account the 
influence of the Yukawa couplings on the evolution of gauge couplings.
We use below the RG equations for  Yukawa couplings that can be derived using 
ref.\cite{9,10}.
First, we evolute the RG equaions from 
$m_Z$ to $m_{top}$ using the SM $\beta$-functions
and then  from $m_{top}$ to $\Lambda_m$ 
(taken as 100TeV) using the MSSM ones.
 Above $\Lambda_m$ we take  into account  the effects of the messenger fields:
\begin{eqnarray}
\mu \frac{d g_1}{d\mu} &=& \frac{1}{16\pi^2} \frac{38}{5} g_1^3 + \left(
\frac{1}{16\pi^2} \right)^2 g_1^3 \left( \frac{632}{75}g_1^2 +\frac{36}{5}
g_2^2 +\frac{296}{15} g_3^2 -\frac{4}{5} Y_D^2 -\frac{6}{5}Y_L^2 
-\frac{26}{5} Y_t^2 -\frac{14}{5}Y_b^2 -\frac{18}{5} Y_{\tau}^2 \right)\nonumber\\
\mu \frac{d g_2}{d\mu} &=& \frac{1}{16\pi^2} 2 g_2^3 + \left(
\frac{1}{16\pi^2} \right)^2 g_2^3 \left( \frac{12}{5}g_1^2 +32
g_2^2 +24 g_3^2 -2 Y_L^2 -6 Y_t^2 -6 Y_b^2 -2 Y_{\tau}^2 \right)\nonumber\\
\mu \frac{d g_3}{d\mu} &=& -\frac{1}{16\pi^2} 2 g_3^3 + \left(
\frac{1}{16\pi^2} \right)^2 g_3^3 \left( \frac{37}{15}g_1^2 +9
g_2^2 +\frac{76}{3} g_3^2 -2 Y_D^2 -4 Y_t^2 -4 Y_b^2  \right)
\end{eqnarray}
The evolution of Yukawa couplings themselves are taken into account in one loop 
approximation and are  carried out using the following RG equations
(for scales larger than $\Lambda_m$):
\begin{eqnarray}
\mu \frac{d Y_D }{d\mu} &=& \frac{Y_D}{16\pi^2} \left( -\frac{4}{15}g_1^2 -\frac{16}{3} 
g_3^2 +5 Y_D^2 +2 Y_L^2\right)\nonumber\\
\mu \frac{d Y_L }{d\mu} &=& \frac{Y_L}{16\pi^2} \left( -\frac{6}{5}g_1^2 -6
g_2^2 +6 Y_D^2 +8 Y_L^2\right)\nonumber\\
\mu \frac{d Y_t }{d\mu} &=&\displaystyle\frac{Y_t}{16\pi^2}
\left(-\displaystyle\frac{13}{15}g_1^2-3g_2^2-\displaystyle\frac{16}{3}
g_3^2+6Y_t^2+Y_b^2\right)\nonumber\\
\mu \frac{d Y_b }{d\mu} &=&\displaystyle\frac{Y_b}{16\pi^2}
\left(-\displaystyle\frac{7}{15}g_1^2-3g_2^2-\displaystyle\frac{16}{3}g_3^2
+Y_t^2+6Y_b^2+Y_\tau^2\right)\nonumber\\
\mu \frac{d Y_\tau }{d\mu} &=&\displaystyle\frac{Y_\tau}{16\pi^2}
\left(-\displaystyle\frac{9}{5}g_1^2-3g_2^2+3Y_b^2+4Y_\tau^2\right)
\end{eqnarray}
Our notations are self-evident:
$Y_t, Y_b$ and $Y_\tau$ are Yukawa couplings of $top$, $bottom$, and $\tau$ 
to the corresponding Higgs doublets
in the miminal SU(5) SUSY  model.
$Y_D$ and $Y_L$ are the messenger Yukawa coupling constants to the
 singlet S.
We proceed then in  the same way  as in ref. \cite{8}.
We use the evolution of $\alpha_1$ and $\alpha_2$ to
determine the GUT scale $M_{GUT}$ where  $\alpha_1$ and $\alpha_2$
unify into $\alpha_5$,  and then get the mismatch between
$\alpha_5 (M_{GUT})$ and $\alpha_3 (M_{GUT})$, 
which  should be  attributed to the GUT  threshold effects, that are 
assumed to originate from the color triplet $H_c$.
We get  the mass of the color-triplet Higgs $M_{H_C}$ from 
its dependence on the threshold effects 
(the detailed formulae can be found in \cite{8}).
Once $M_{H_C}$ is determined, we use the standard formulae for the proton
lifetime(see ref. \cite{9}) to find the partial width and to compare them
with the current experimental bounds.

In our numerical evolution,
we use the following boundary conditions.
First,
we made a standard choice of the masses of $b$-quark , $\tau$-lepton
and the top-quark as: $m_b=4.5$ GeV, $m_{\tau}=1.78$ GeV and  $ m_t=175$ GeV.
We do not depict uncertainties in these boundary conditions
due to  masses, since their influence on the results is not large.
For the messenger Yukawa couplings we take the largest possible
values :
\begin{eqnarray}
Y_D(\mu=\Lambda_m)&=&0.9,\nonumber\\
Y_L(\mu=\Lambda_m)&=&0.452.
\label{7}
\end{eqnarray}
If we choose the boundary conditions to be given by eq. (\ref{7}), $Y_D$ and
 $Y_L$  unify into a common value
at the GUT scale.
If  inputs larger than those in (\ref{7})
  is taken, the messenger Yukawa couplings
will blow up at the GUT scale.
The boundary conditions for gauge couplings
 are determined from LEP data\cite{11}:
\begin{eqnarray}
\alpha_1^{-1}(m_Z)&=&58.96\pm 0.05,\nonumber\\
\alpha_2^{-1}(m_Z)&=&29.63\pm 0.05.
\end{eqnarray}
 For the QCD coupling we use $\alpha_3(m_Z)=0.116\pm 0.005$ \cite{11,8} -----
the number that incorporates both low energy data and 
LEP1 data. All the gauge couplings are taken within $2\sigma$ variations
in our estimations.

 \section{The Results.}

\par Our main results are depicted in Figs. 1-6.
First, we depict the mass of the color-triplet,  $M_{H_C}$,
as a function of $\alpha_3(m_Z)$ for different $\tan \beta$.
These results are depicted in Fig. 1-3 for different values
of $\alpha_1(m_Z)$ and $\alpha_2(m_Z)$.
We take the lowest bound of $\tan \beta$
as 0.85. Below this bound  the Yukawa couplings will
blow up at the GUT scale. 
This lowest bound can be achieved
only when the Yukawa couplings are included in the
RG analyses.
We see that for reasonable values of  $\tan \beta$,
Yukawa couplings of messenger quarks and leptons actually 
do not influence $M_{H_C}$ greatly.
However, when we  consider  the different values of $\alpha_1$ and $\alpha_2$
within $2\sigma$ variations, we see that $M_{H_C}$ 
can change quite significantly
as a function of $\alpha_1(m_Z)$ and $\alpha_2(m_Z)$.
The most favored  $M_{H_C}$ is gained
when $\alpha_1(m_Z)$ is taken at its upper bound while
$\alpha_2(m_Z)$ the lower one.

Next, we consider the   decay rate of  a typical process
$n\rightarrow K^0\bar \nu$.
This mode was argued in ref. \cite{8} to be the most
appropriate one in analysing proton decay
bound in  SUSY GUT. 
The results are shown in Figs. 4-6.
We see that the allowed region of $\alpha_3$ and  $\tan \beta$
depends on the inputs $\alpha_1(m_Z)$ and $\alpha_2(m_Z)$.
Inclusion of the 
usual Yukawa couplings 
in RG equations improves the situation with the
proton lifetime.
On the other hand, inclusion of the 
messenger Yukawa couplings  even
worses  the situation slightly.
As the current experimental bound\cite{pdg} of proton decay lifetime is 
concerned,
the only   window  where the model survives 
is the region of small  $\tan \beta$
combined with  large $\alpha_1(m_Z)$
and small $\alpha_2(m_Z)$ as input.

Our general conclusions are that, first, the inclusion of Yukawa couplings
does not lead to significant changes in the mass of color triplet and
the allowed parameter space for the model we consider. Second,
the allowed parameter space, color triplet mass and proton lifetime 
seem to be quite sensitive to exact values of $\alpha_1$ and $\alpha_2$
at $M_Z$ scale.
\section*{Acknowledgments}

The authors thank M. Shifman for useful discussion.

\section*{Figure Captions}
\noindent

Fig. 1 The dependence of the color triplet boson mass on $\tan \beta$
and $\alpha_3(m_Z)$, when $\alpha_1(m_Z)$ and $\alpha_2(m_Z)$ are both taken 
as central values of experiment results. 
The dotted and the dash-dotted lines
correspond to the results without
any Yukawa couplings and with only messenger Yukawa couplings, respectively.
The dashed and  solid 
lines correspond to results 
with  all Yukawa couplings included for $\tan \beta =0.85$ and
$\tan \beta =50$, respectively. 

Fig. 2 Same as Fig.1
 except that both $\alpha_1(m_Z)$ and $\alpha_2(m_Z)$ are  taken 
$2\sigma$ smaller
 than the experimental central values.

Fig. 3 Same as Fig.1 except that $\alpha_1(m_Z)$ is taken $2\sigma$ larger and 
$\alpha_2(m_Z)$ is taken $2\sigma$ smaller than the  experimental 
central values. 

Fig. 4 $\alpha_3(m_Z)$ and $\tan \beta$ plane excluded by the proton decay
constraint,
when $\alpha_1(m_Z)$ and $\alpha_2(m_Z)$ are both taken their central values.
 The dash-dotted, dashed and solid lines correspond 
to the cases without
Yukawa couplings, with only messenger Yukawa couplings, and
with  all Yukawa couplings included, respectively. 

Fig. 5 Same as Fig.4 except that
both of  $\alpha_1(m_Z)$ and $\alpha_2(m_Z)$ are  taken $2\sigma$ smaller
than their central values.

Fig. 6 Same as Fig.4 except that
 $\alpha_1(m_Z)$ is taken $2\sigma$ larger and $\alpha_2(m_Z)$ is taken 
$2\sigma$ smaller than their central values.

\newpage
\begin{figure}
\centerline{\epsffile{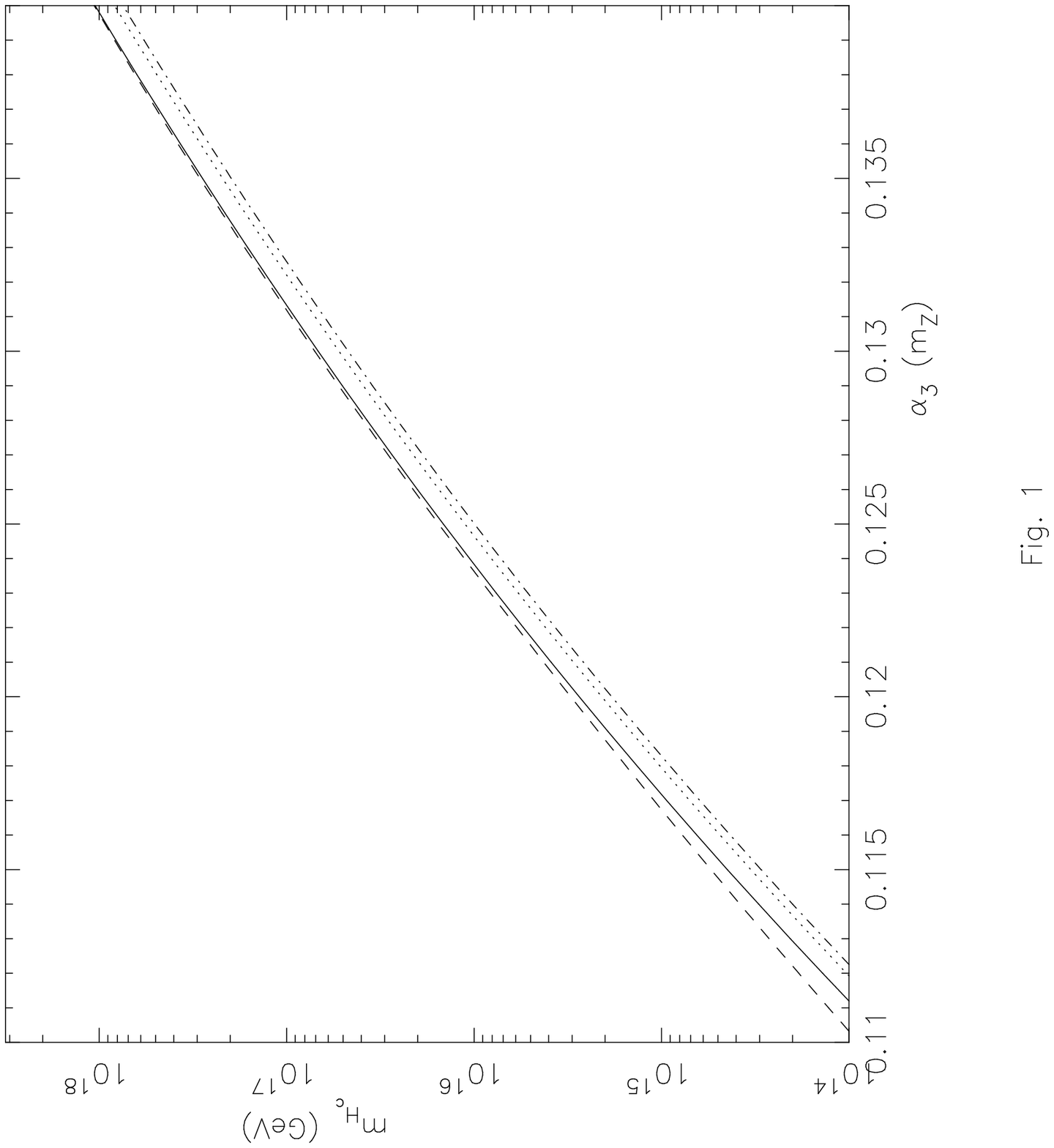}}
\end{figure}

\begin{figure}
\centerline{\epsffile{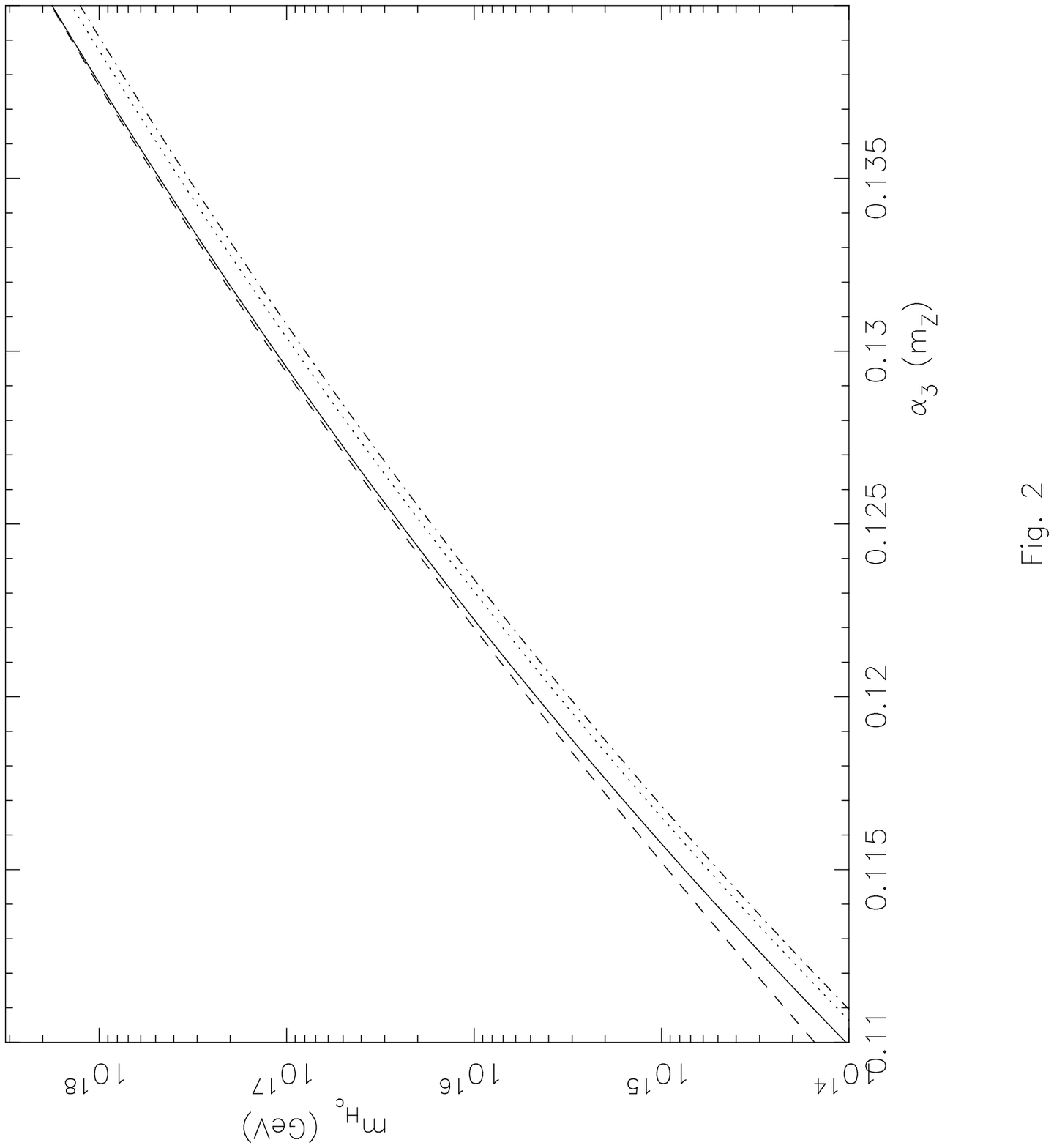}}
\end{figure}

\begin{figure}
\centerline{\epsffile{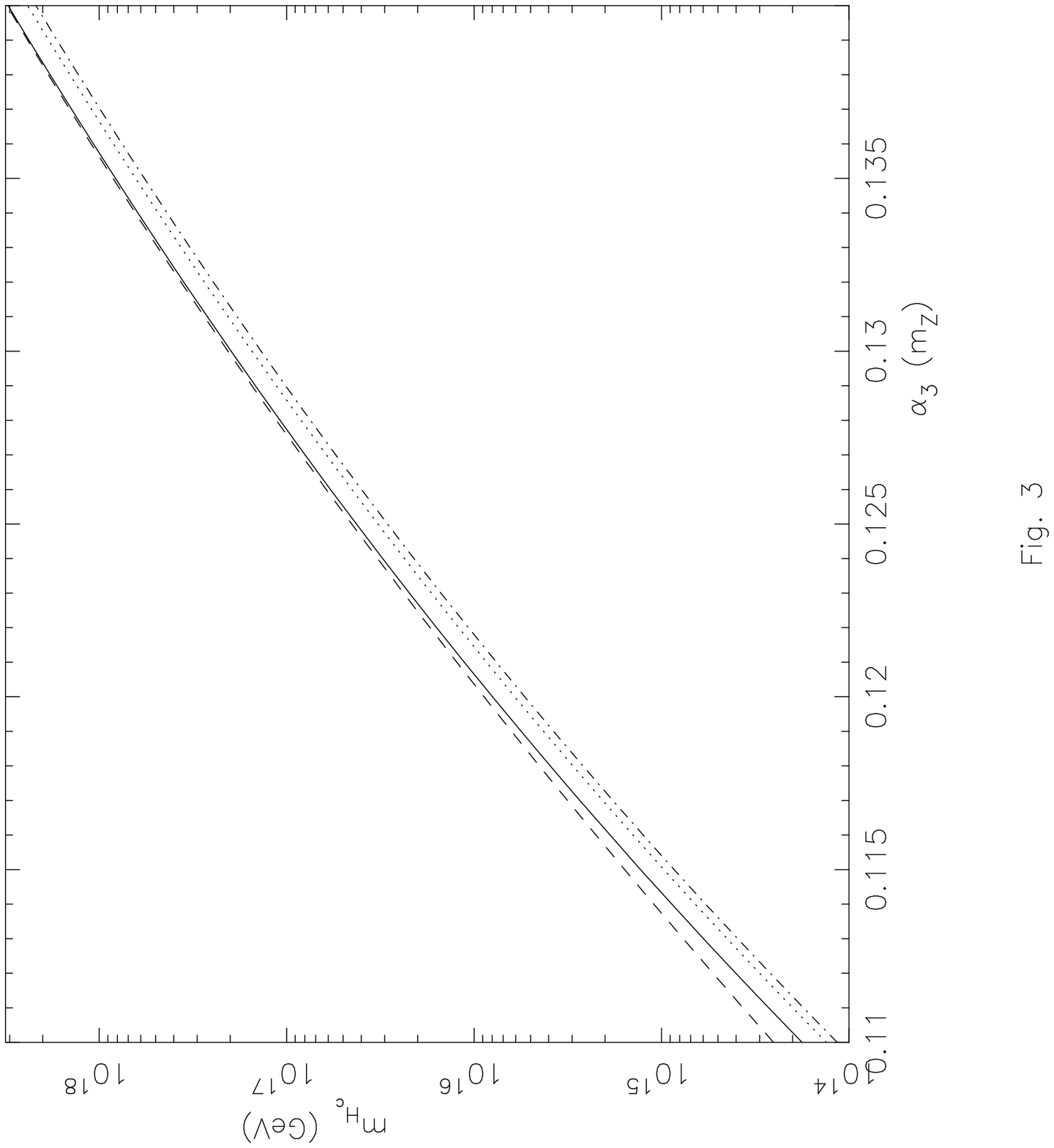}}
\end{figure}

\begin{figure}
\centerline{\epsffile{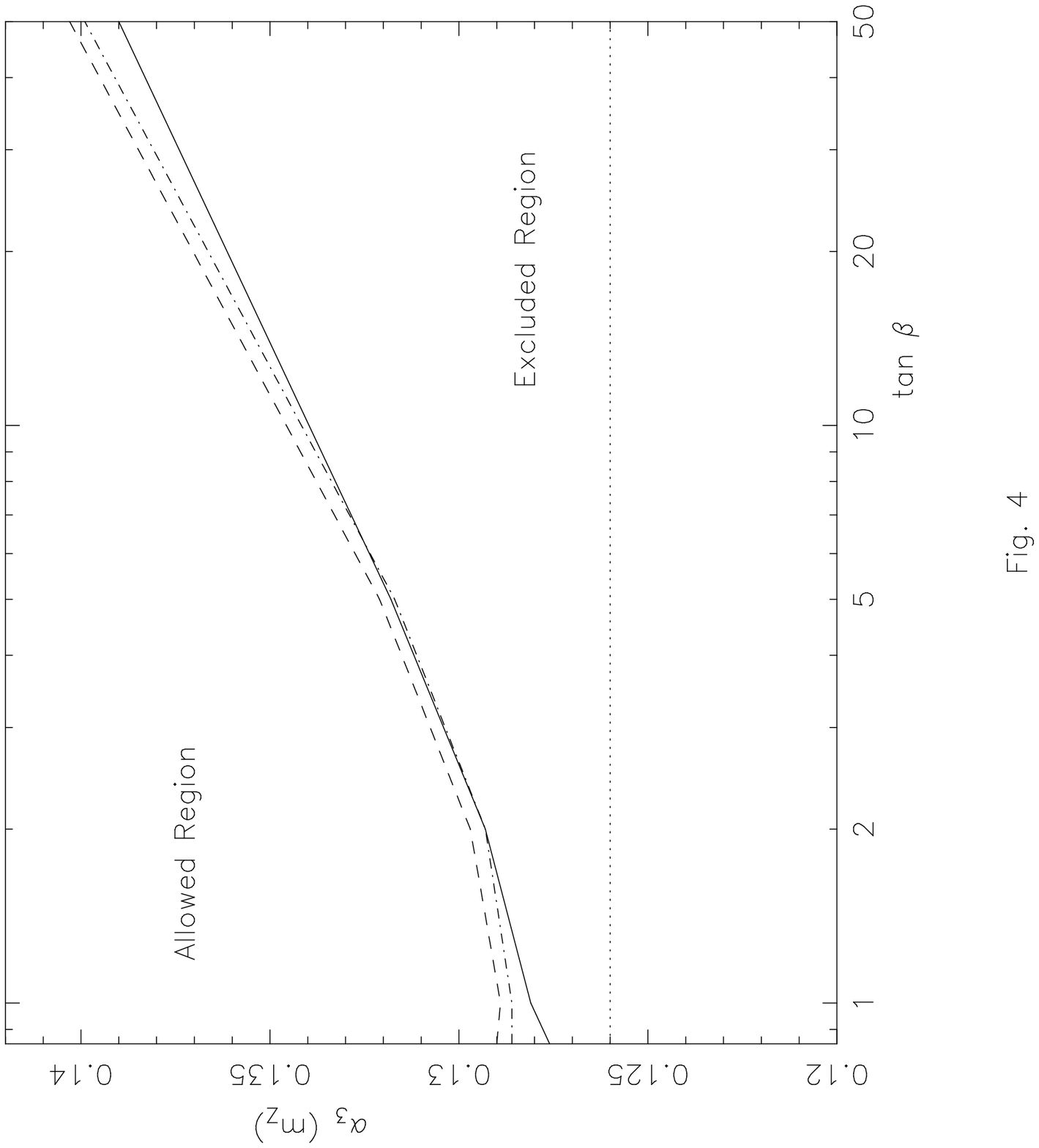}}
\end{figure}

\begin{figure}
\centerline{\epsffile{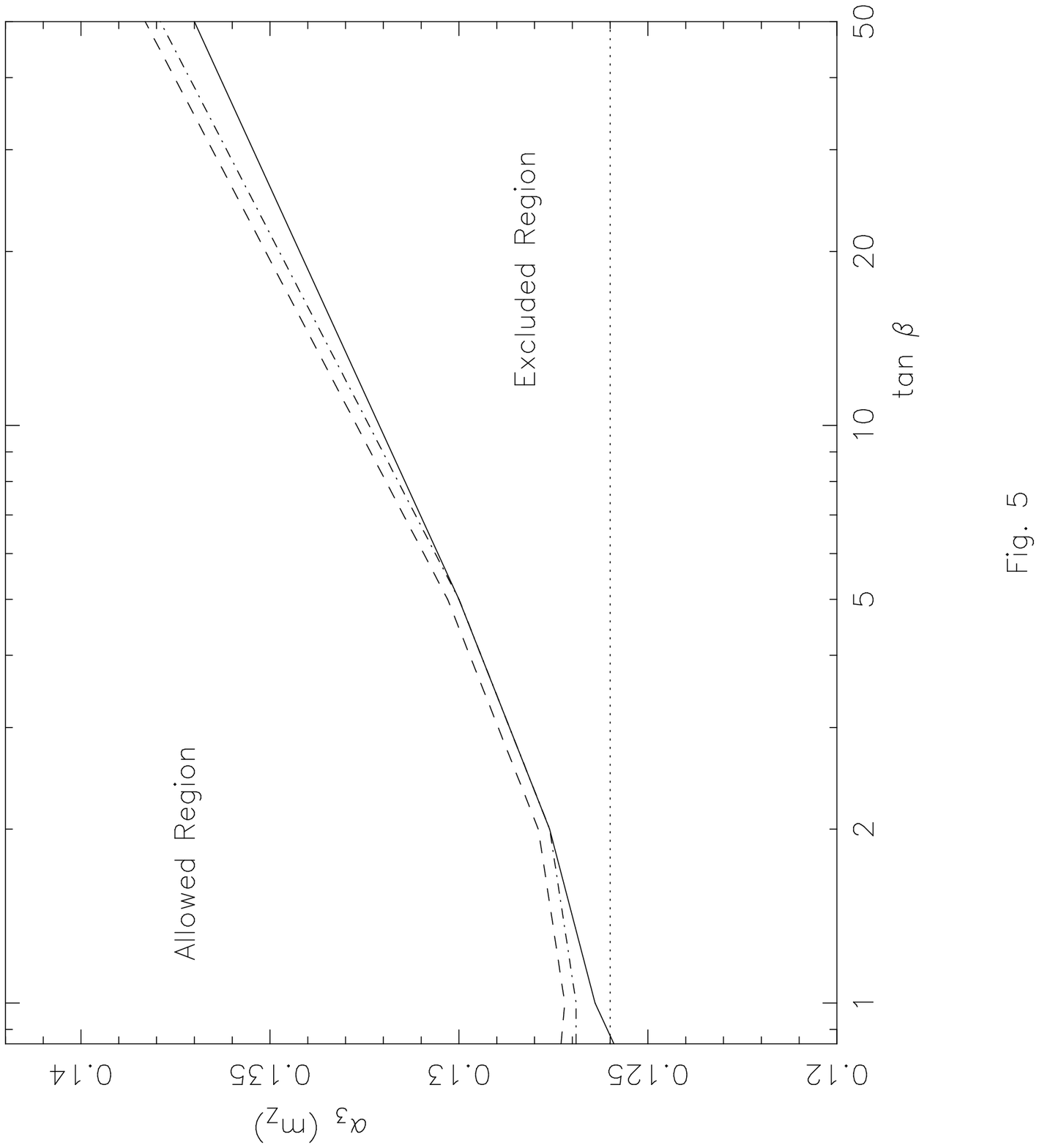}}
\end{figure}

\begin{figure}
\centerline{\epsffile{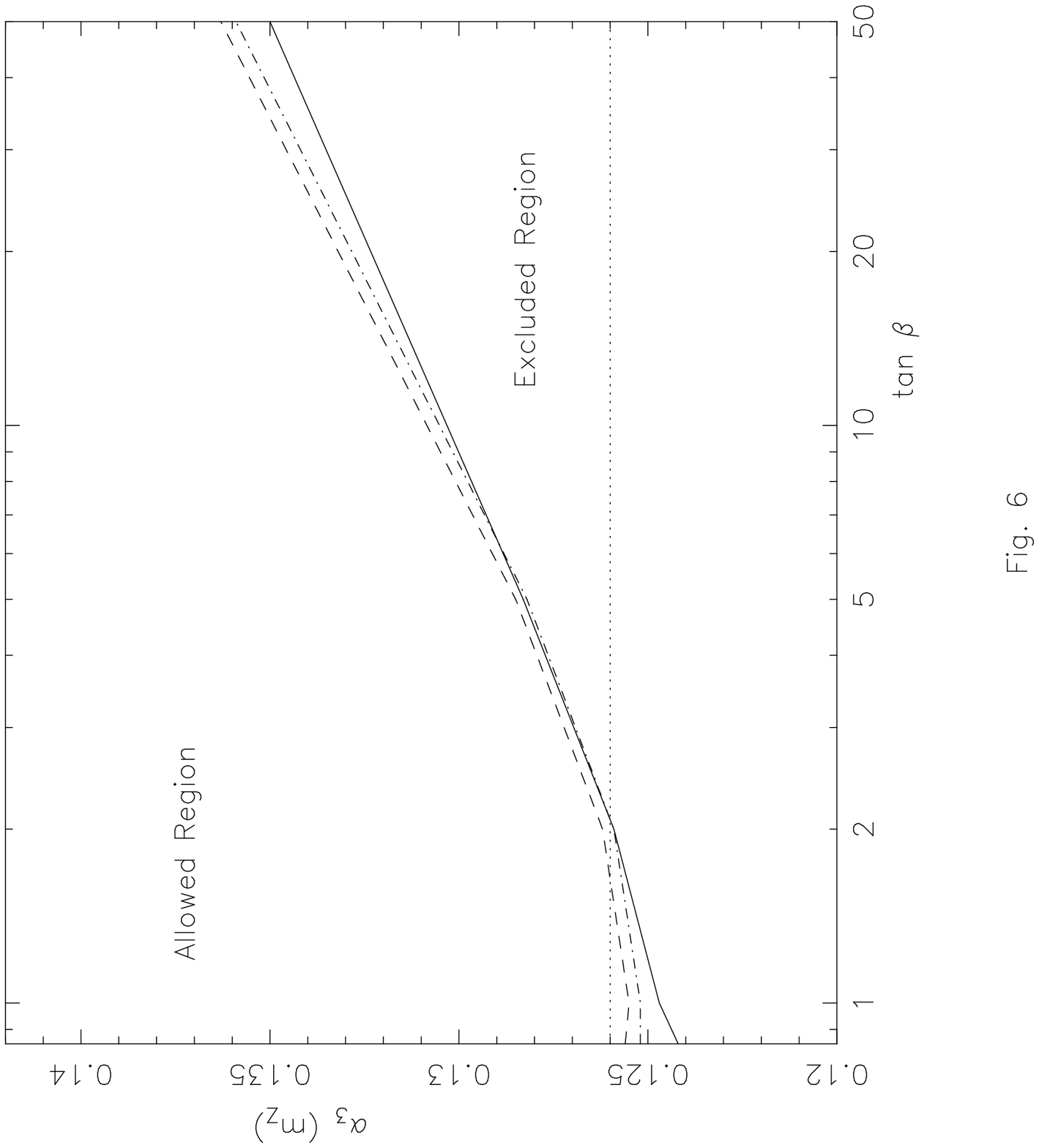}}
\end{figure}

\end{document}